\documentclass[12pt]{article}
\usepackage{amssymb}

\usepackage{graphicx}
\usepackage{amsmath}

\input{tcilatex}

\begin{document}

\title{Beginnings of the Cauchy problem.}
\author{Yvonne Choquet-Bruhat}
\maketitle

\section{Introduction.}

I was asked to write a short article on the early works on the Cauchy
problem for the Einstein equations, in honor of the hundredth anniversary of
General Relativity. I accepted with pleasure, but I realized when I started
to work on this project that it was more difficult than I thought. I have
never been really interested in who did something first, and in fact it is
often difficult to ascertain. Ideas have almost always one or several
preliminaries, attribution of a name to the final flower is therefore
somewhat arbitrary. The work of a true historian is long and difficult, I am
not an historian. I often quote, not the first note touching a subject, but
a later paper more complete and easier to find. The shortness of this
article does not enable me to enter into details. Of course what I know best
is my own work, it it is part of my excuse for often quoting it. I apologize
to all live or dead authors to whom I did not make enough deserved
references. Other sources of information, including my own articles, can
compensate my shortcomings.

\section{Preliminary definitions}

The \textbf{Einstein equations} are a geometric system for a pair $(V,g)$,
with $V$ an $n+1$ dimensional differentiable manifold, $n=3$ in the
classical case, and $g$ a pseudo-Riemannan metric of Lorentzian signature.
In vacuum they express the vanishing of the Ricci tensor 
\begin{equation*}
\text{Ricci}(g)=0,
\end{equation*}
equivalently the vanishing of the Einstein tensor ($R(g)$ is the scalar
curvature of $g)$%
\begin{equation*}
\text{Einstein}(g):=\text{Ricci}(g)-\frac{1}{2}gR(g)=0.
\end{equation*}
These equations are invariant under diffeomorphisms of $V$ and associated
isometries of $g$.

The Bianchi identities satisfied by the Riemann tensor imply, by two
contractions, identities for the Einstein tensor which read\footnote{%
We denote $S$ the Einstein tensor. It is denoted $G$ by some authors.} in
local coordinates $x^{\alpha },$ $\alpha =0,1,...n,$%
\begin{equation*}
\nabla _{\alpha }S^{\alpha \beta }\equiv \nabla _{\alpha }(R^{\alpha \beta }-%
\frac{1}{2}g^{\alpha \beta }R)\equiv 0,
\end{equation*}
where $\nabla $ is the covariant derivative in the metric $g$

The vacuum Einstein equations constitute, from the analyst's point of view,
a system of $\frac{(n+1)(n+2)}{2}$\ second order quasilinear\footnote{%
i.e. linear with respect to second derivatives} partial differential
equations for the $\frac{(n+1)(n+2)}{2},$ $10$ in the classical case $n=3$,
coefficients $g_{\alpha \beta }$ of the metric $g$ in local coordinates.
However these equations are not independent because of the above identities.

The \textbf{Cauchy problem} for a system of $N$ second order quasilinear
partial differential equations with unkown $u$ a set of $N$ functions $%
u_{I}, $ $I=1,...,N$ on $R^{n+1},$ 
\begin{equation*}
A_{J}^{I,\alpha \beta }(u,\partial u)\partial _{\alpha \beta
}^{2}u_{I}=f_{J}(u,\partial u),\text{ \ \ }\partial _{\alpha }:=\frac{%
\partial }{\partial x^{\alpha }},
\end{equation*}
is the search for a solution $u$\ which takes, together with its set $%
\partial u$ of first order partial derivatives, given values $\bar{u},$ $%
\overline{\partial u}$ on a given $n$ dimensional submanifold $M$. The
elements of the characteristic determinant of this system, for a function $u$
at a point $x,$ are the second order polynomials in a vector $X:$ 
\begin{equation*}
D_{J}^{I}(u,\partial u,X):=A_{J}^{I,\alpha \beta }(u,\partial u)X_{\alpha
}X_{\beta },\text{ \ \ }
\end{equation*}
A submanifold with equation 
\begin{equation*}
\phi (x^{\alpha })=0
\end{equation*}
is called characteristic at a point for the considered system and initial
values $\bar{u},$ $\overline{\partial u}$ if the determinant with elements $%
(D_{J}^{I})(\bar{u},\overline{\partial u},\partial \phi ),$ polynomial of
order $2N,$ vanishes at that point. The Cauchy-Kovalevski theorem says that
if this system has analytic coefficients the Cauchy problem with analytic
given initial data has one and only one analytic solution in a neighbourhood
of $M$ if this sumanifold is everywere non characteristic.

The Cauchy-Kowalevski theorem does not apply directly to the Einstein
equations: their characteristic determinant is identically zero for any
metric as can be foreseen from the identities satisfied by the Enstein
tensor. The Cauchy problem for the Einstein equations is non standard and
has led to interesting and difficult works.

\section{Analytic results.}

Hilbert by Lagrangian methods and Einstein himself by approximation studies
had been interested in what Einstein called ''the force'' of his equations,
that is the generality of their solutions. However the history of exact
results on the general Cauchy problem for the Einstein equations starts only
in 1927 with the 47 pages book ''Les \'{e}quations de la gravitation
Einsteinienne'' by Georges Darmois, a professor of mathematics in the
University of Paris\footnote{%
In those times there was only one university of Paris. Sciences, letters,
law and arts were housed in a building called the Sorbonne, G. Darmois was a
man of varied interests. In 1948 he taught a course on probabilities which I
attended. He was a member of the french Academy in the section ''Astronomy''.%
}. Darmois considers (case $n=3)$ a submanifold $M$ with equation $x^{0}=0$
and data on $M$ functions of the $x^{i},$ $i=1,2,3,$ which will be the
values on $M$ of $g_{\alpha \beta }$ and $\partial _{0}g_{\alpha \beta }.$
The values on $M$ of the first and second partial derivatives $\partial
_{\lambda \mu }^{2}g_{a\beta }$ are then determined by derivation of the
data on $M$ except for the second transversal derivatives $\partial
_{00}^{2}g_{a\beta }.$ Darmois finds by straightforward computation the
identities: 
\begin{equation}
R_{ij}\equiv -\frac{1}{2}g^{00}\partial _{00}^{2}g_{ij}+f_{ij}(g_{\alpha
\beta },\partial _{\lambda }g_{a\beta },\partial _{\lambda h}^{2}g_{a\beta })
\end{equation}
\begin{equation*}
R_{i0}\equiv \frac{1}{2}g^{j0}\partial _{00}^{2}g_{ij}+f_{io}(g_{\alpha
\beta },\partial _{\lambda }g_{a\beta },\partial _{\lambda h}^{2}g_{a\beta })
\end{equation*}
\begin{equation*}
R_{00}\equiv -\frac{1}{2}g^{ij}\partial _{00}^{2}g_{ij}+f_{00}(g_{\alpha
\beta },\partial _{\lambda }g_{a\beta },\partial _{\lambda h}^{2}g_{a\beta }
\end{equation*}
The derivatives $\partial _{00}^{2}g_{\alpha 0}$ do not appear in any of
these equations, as Darmois already foresaw because a change of coordinates
preserving $M$ pointwise does not change $\partial _{00}^{2}g_{ij}$ on $M,$
but permits to give arbitrary values to $\partial _{00}^{2}g_{\alpha 0}.$

If $g$ satisfies the vacuum Einstein equations, the equations $R_{ij}=0$
determine $\partial _{00}^{2}g_{ij}$ on $M$ if $g^{00}$ does not vanish
there. Darmois concludes that significant discontinuities of the second
derivatives of the gravitational potentials can occur across the submanifold 
$M,$ $\phi (x^{\alpha })\equiv x^{0}=0,$ only if $g^{00}=0$ on $M,$ that is
if the hypersurface $M$ is tangent to the null cone of the Lorentzian metric 
$g,$ whose normals in the cotangent space satisfy the equation $g^{\alpha
\beta }\partial _{\alpha }\phi \partial _{\beta }\phi =0.$ This result,
though not a proof of it, is in agreement with the propagation of
gravitation with the speed of light, fact already deduced by Einstein from
approximations.

Darmois continues his study by remarking that, if $g^{00}\not=0,$ it is
possible to extract $\partial _{00}^{2}g_{ij}$ from the equation $R_{ij}=0$
and, replacing these in the other equations by the so calculated
expressions, obtain four equations which depend only on the initial data,
equations which we now call the constraints; he mentions that they are the
Gauss - Codazzi equations known from geometers and indicates that a solution
of the equations $R_{ij}=0$ with data satisfying the constraints will
satisfy the whole set, at least in the analytic case, due to the contracted
Bianchi identities. Darmois recognizes that an analyticity hypothesis is
physically unsatisfactory, because it hides the propagation properties of
the gravitational field.

Darmois also studies, again in the analytic case, what initial data to give
on a characteristic hypersurface $S_{0}$. He shows that they are the trace
of the spacetime metric on the hypersurfacc and proves, in the analytic
case, the existence of a local solution to the vacuum Einstein equation
which is uniquely determined if one gives also its value on a 3-dimensional
manifold $T$ transversal to $S_{0}$ or, what is equivalent for analytic
functions, the values of all its derivatives at points of the intersection
of $T$ and $S_{0}$. To show this, \ he uses adapted coordinates to decompose
the problem into an evolution of some components of the metric to satisfy
part of the Einstein equations, and the Bianchi identities to show that the
remaining equation is also satisfied\textrm{. }The method used by Darmois
does not extend to the non analytic case, though the introduction of the
data of the trace of the metric on a second hypersurface, transversal to the
characteristic one, but also characteristic in the non analytic case, has
been successfully used, in particular in the nineties by Rendall. Before
that, inspired by the general theorems of Leray for data with support
''compact towards the past ''\footnote{%
That is itersected along a compact set by the past of any point.}, the non
analytic Cauchy problem was treated for data supported by a characteristic
cono\"{i}d\footnote{%
Y.\ Bruhat \textit{''Probl\`{e}me des conditions initiales sur un
cono\"{i}de caract\'{e}ristique} C.R.\ Acad. Sci \textbf{256}, 371-373, 1963.
\par
F. Cagnac ''\textit{Probl\`{e}me de Cauchy sur les hypersurfaces
caract\'{e}ristiques des \'{e}qutions d'Einstein du vide''} C.R. Acad Sci 
\textbf{262} 1966
\par
For more recent works see papers by Cagnac and his students, in particular
Dossa. Still more recent, Choquet-Bruhat, Chrusciel and Martin-Garcia, also
Chrusciel and collaborators}.

In the remainder of his book, after quoting the works of Droste and
Schwarzchild on the solution with spherical symmetry, Darmois studies
solutions with axial symmetry\footnote{%
For other early work see J. Delsarte ''\textit{Sur les ds}$^{2}$\textit{\
d'Einstein \`{a} sym\'{e}trie axiale' }Hermann 1934.}.

Darmois had mentioned the geometric character of the constraints but he had
worked in special coordinates, namely in Gaussian coordinates; that is, with
timelines geodesics normal to the initial manifold, the quantities that we
call now lapse and shift are then equal respectively to one and zero.
Lichnerowicz, a bright student of the Ecole Normale Sup\'{e}rieure, had
asked from Elie Cartan a subject for his thesis, and Cartan had proposed the
proof of a conjecture he had on a property of symmetric spaces that himself
had not been able to prove for some time. Lichnerowicz proved it thirty
years later, but when he met Darmois by chance in 1937 he was rather
discouraged, and followed the suggestion of Darmois to work instead of the
too difficult problem proposed by Cartan to problems on mathematical
relativity which were many and little considered at the time. Lichnerowicz
who was a man of varied interests, from algebra and differential geometry to
theoretical physics and philosophy, followed Darmois suggestion and
completed quickly a thesis which appeared as a book\footnote{%
Lichnerowicz A. ''probl\`{e}mes globaux en m\'{e}canique relativiste'',
Hermann 1939}. In this book the Darmois computations on the Cauchy problem
are extended to the case of a non constant lapse but the shift is kept zero.
Lichnerowicz proposed the extension of the 3+1 decomposition to a non zero
shift to one of his two first students\footnote{%
The other wasYves Thiry who worked on geometrical aspects and physical
interpretation of the five dimensional unitary theory of Jordan, extension
of Kaluza and Klein work.}, the author of this article, who did it through
the \ use of the Cartan calculus in orthonormal frames, giving thus the
general geometric formulas of the $n+1$ decomposition of the Ricci and
Einstein tensor on a sliced manifold $M\times R$ in terms of the geometric
elements: $t$ dependent induced metric and extrinsic curvature of the slices 
$M\times \{t\}$. This led to a preliminary publication\footnote{%
Four\`{e}s-Bruhat C. R. Acad Sci. Paris 1948}. However this formulation did
not lead to a new existence theorem for the solution of the evolutionary
Cauchy problem and the full detailed article was written only later\footnote{%
Y. Four\`{e}s-Bruhat ''\textit{sur l'int\'{e}gration des \'{e}quations de la
Relativity G\'{e}n\'{e}rale}'' J. Rat. Mech. and Anal. \textbf{5} 951-966
1956}.

Lichnerowicz, as a student of Elie Cartan, had a formation of geometer. He
insisted on geometric formulations and started the study of global problems%
\footnote{%
A. Lichnerowicz \textit{''Probl\`{e}mes globaux en m\'{e}canique relativiste}%
''\ Hermann et Cie, 1939 and A.\ Lichnerowicz ''\textit{Th\'{e}ories
relativistes de la gravitation et de l'\'{e}lectromagn\'{e}tisme''} Masson
1955 which contains also a study of the 5 dimensional and the non symmetric
unitary theories.}. He stated two what he called\footnote{%
Better named ''conjectures''.} ''propositions'', A and B, for the
Einsteinian spacetimes which he called regular (The metric had to be $C^{2}$
by pieces with first derivatives satisfying ''junction conditions'')%
\footnote{%
The relevant condition is in fact that the Einstein equations are satisfied
in a generalized sense: see YCB ''Espaces Einsteiniens g\'{e}n\'{e}raux,
chocs gravitationnels''\ Ann. Inst. Poincar\'{e}, 8 n$^{0}4$, 327-338, 1968.}%
, A: introduction of matter sources in a vacuum spacetime can be done only
in domains where this spacetime has singularities; B: the only complete
vacuum spacetime with compact or asymptotically Euclidean space sections is
flat. He proved B in the case of stationary\footnote{%
That is invariant under a timelike one parameter isometry group. The static
case (timelines orthogonal to space sections) had been proved earlier by
Racine, C. R. Acad. Sciences 192, 1533 1931., another student of Darmois}
spacetimes; that is, the non existence of gravitational solitons. The
Lichnerowicz result was much appreciated by Einstein who believed for
physical reason that any complete asymptotically Euclidean vacuum
Einsteinian spacetime should be Minkowski\footnote{%
The Christodoulou-Klainerman global existence theorem (1989) has proven that
the conjecture was false without stronger hypothesis than the ones
originally made on the decay at infinity; that is, vanishing of the ADM mass
as shown by Shoen and Yau.}.

\section{Non analytic local existence, causality and gravitational waves.}

It had already been stressed by Darmois on the one hand that analyticity was
a bad physical hypothesis, on the other hand that a choice of coordinates
was necessary to construct solutions of the Cauchy problem. The problem had
interested Einstein himself and already\footnote{%
Einstein A. Sitzgsb 1918.} in 1918 he had used coordinates satisfying the
flat spacetime wave equation to construct approximated solutions of the
vacuum Einstein equations near the Minkowski spacetime.

Darmois\footnote{%
Darmois quotes as sources:
\par
De Donder ''La gravifique Einsteinienne ''Mem. Sci. Math. 1925 Gauthier
Villars
\par
This article can be found on numdam. It uses the Lagrangian formulation of
Einstein equations .
\par
Lanczos K. Physzeitshrift p.137 1922} considers coordinates $x^{\lambda }$
which he calls ''isothermes''; they satisfy the wave equations 
\begin{equation}
\square _{g}x^{\lambda }\equiv g^{\alpha \beta }\nabla _{\alpha }\partial
_{\beta }x^{\lambda }=0;
\end{equation}
that is 
\begin{equation}
F^{\lambda }\equiv g^{\alpha \beta }\Gamma _{\alpha \beta }^{\lambda }=0.
\end{equation}
Such coordinates are now called ''harmonic'' by analogy with solutions of
the Laplace equations, or ''wave'' as suggested by Klainerman as being more
appropriate.

By a straightforward concise and precise computation Darmois obtains the
decomposition of the Ricci tensor of a pseudo Riemannian general metric as
the sum of a second order system for the components of the metric and a term
which vanishes identically in harmonic coordinates 
\begin{equation}
R_{\alpha \beta }\equiv R_{\alpha \beta }^{(h)}+L_{\alpha \beta },\text{ \
with \ }L_{\alpha \beta }\equiv {\frac{1}{2}}\{g_{\alpha \lambda }\partial
_{\beta }F^{\lambda }+g_{\beta \lambda }\partial _{\alpha }F^{\lambda }\}.
\end{equation}
where 
\begin{equation}
R_{\alpha \beta }^{(h)}\equiv -\frac{1}{2}g^{\lambda \mu }\partial _{\lambda
\mu }^{2}g_{\alpha \beta }+P_{\alpha \beta }(g,\partial g).
\end{equation}
with $P$ a quadratic form in the components of $\partial g$ whose
coefficients are polynomials in the components of $g$ and its contravariant
associate.

In harmonic coordinates the Einstein equations in vacuum reduce to the
quasilinear quasi diagonal second order system $R_{\alpha \beta }^{(h)}=0.$

In the years shortly before the second world war great names in mathematics
were working on the Cauchy problem for a second order equation of the type
then called ''hyperbolic normal'', that is principal coefficients of
Lorentzian signature. The Hadamard method of parametrix for solution of
linear equations seemed difficult to use for non linear equations. On the
other hand a new method, energy estimates, introduced by Friedrichs and
Lewy, was a subject of active research. An application of the energy
estimates to the reduced vacuum Einstein equations enabled Stellmacher%
\footnote{%
K. Stellmacher. Math. Annalen 115, 1938.} to prove an uniqueness theorem for
a local solution of the Cauchy problem for the reduced equations, with
domain of dependence determined by the light cone; that is, a causality
property. However Stellmacher did not prove an existence theorem, in spite
of a paper of \ Schauder\footnote{%
Schauder J. Fundamenta mathematicae, 24 1935, p213-246.
\par
Schauder, a collaborator and friend of J. Leray was a victim of the
holocaust, having refused to follow Leray advice to leave Germany when it
was still possible for jews.} where was sketched an existence proof for a
solution of one quasilinear second order equation equation by using energy
estimates.

I was encouraged to look for the solution of the non analytic Cauchy problem
for the Einstein equations in 1947 by Jean Leray who was giving a series of
lectures on Cartan exterior differential systems, of which I was one of the
few attendants. Leray gave me the name of Schauder as a reference but I
found only his paper on hyperbolic system in two variables\footnote{%
J.\ Schauder Comm. Math. Helv.\textbf{9} 1936-1937.} written later, which I
tried somewhat painfully to read, knowing no german. By chance I fell on a
paper by Sobolev\footnote{%
S.\ Sobolev \ \textit{''Methode nouvelle \`{a} r\'{e}soudre le probl\`{e}me
de Cauchy pour les \'{e}quations lin\'{e}aires hyperboliques normales''}
Rec. Math. Moscou N. s. 1936.}, in french, which gives a construction of an
elementary kernel for a second order linear hyperbolic equation in dimension
3+1 without to have to resort to a ''finite part''\ parametrix nor to the
method of descent for the case of even spacetime dimension, as did Hadamard.
The Sobolev parametrix, whose definition extends to quasi diagonal second
order systems, is constructed by solution of a system of integral equations
on the characteristic cono\"{i}d. These equations, together with those
defining the characteristic cono\"{i}d can be used to prove the existence of
a solution of the Cauchy problem for the quasilinear reduced vacuum Einstein
equations in a space of smooth functions\footnote{%
See Four\`{e}s-Bruhat Y. ''Th\'{e}or\`{e}mes d'existence pour certains
syst\`{e}mes d'\'{e}quations aux d\'{e}riv\'{e}es partielles non
lin\'{e}aires''\ Acta Mathematica \textbf{88, 42-225,}\ 1952 and references
therein.
\par
{}}. I showed that the obtained solution satisfies the full Einstein
equations if the initial data satisfy the constraints and that it is locally
geometrically unique\footnote{%
See also Y. Bruhat ''The Cauchy problem''\ in ''Gravitation, an introduction
to current researc''\ Louis Witten ed. Wiley 1962}. This was the subject of
my thesis, its jury included Lichnerowicz, Leray\footnote{%
In fact, while I was still working on my thesis Leray was completing his
momentum work on energy estimates and existence theorems for general
hyperbolic systems, from which I could have deduced the result for the
reduced Einstein equations, but Leray encouraged me warmly to pursue in the
constructive direction I was following in the second order case.} and Marcel
Riesz\footnote{%
Present in Paris at that time. Darmois, an emeritus, could not belong to a
thesis jury.}. I returned later to the construction of the elementary kernel
of a tensorial linear system of second order hyperbolic differential
equations\footnote{%
Y. Choquet-Bruhat \textit{Sur la th\'{e}orie des propagateurs}''\ Annali di
Matematica Serie IV, tomo LXIV -1964.
\par
{}} motivated by works on quantization in curved spacetime by A.
Lichnerowicz who used a propagator, difference of the advanced and retarded
elementary kernels\footnote{%
A. Lichnerowicz ''\textit{Propagateurs et commutateurs en Relatitit\'{e}
G\'{e}n\'{e}rale'', }publications math\'{e}matiques de l'IHES, 1$,$ 1961%
\textit{\ }
\par
See also Bryce DeWitt.\textit{''Quantization of geometry''} in \textit{Les
Houches 1963} Gordon and Breach}. I pointed out that, being obtained by
solving an integral equation on the light cone, the elementary kernel can be
split into the sum of a measure supported by the light cone and a smooth
function in its causal interior sum of a series of ''diffusion terms'',
determined by integrations over characteristic cones with vertices at points
of the previously considered cones\footnote{%
For a lowering of the assumed regularity of the Lorentzian metric see S.
Klainerman and I. Rodnianski ''\textit{The Kirchoff-Sobolev formula''\ }\
Arxiv.math 2006,}. I studied the asymptotic behaviour of these terms.

Einstein, whom I met in 1951 at the Institute for Advanced Study in
Princeton where I was a postdoc at the invitation of J. Leray, made me
explain my thesis on the blackboard of his office; he congratulated me and
invited me to knock at his door whenever I felt like it. I regret to have
done it only a few times, in spite of his always kind welcome. Einstein was
then working with his assistant Bruria Kaufmann on his last unified theory.
His comments were very interesting, but the computations, which himself
enjoyed to do, were quite complicated and the theory rather deceptive%
\footnote{%
Einstein tried at that time to interpret the antisymmetric part of the
second rank tensor as electromagnetism. It appears that this last Einstein
unified theory finds a renewal of interest with another intepretation (see
Damour and Deser)}. I prefered to work at the extension to higher dimensions%
\footnote{%
Y. Four\`{e}s-Bruhat ''\textit{R\'{e}solution du probl\`{e}me de Cauchy pour
des \'{e}quations hyperboliques du second ordre non lin\'{e}aires''} Bull.
Soc. Math. France \textbf{81}, 225-288 1953} of the formulas I had obtained
in spacetime dimension 4 and follow the course of Leray on general
hyperbolic systems. Though I also attended the Oppenheimer seminar on
theoretical physics, where Einstein never came, I did not find there
inspiration for personnal work\footnote{%
My main memory of this seminar is a discussion of quantum vacua and the
intervention of Wigner ''but in vacuum there is nothing, nothing, there can
be only one vacuum'', Wigner, and also Einstein, lived in a time where the
observed world could be thought to obey human scale logic.}.

\section{Equations with sources.}

The existence for classical sources in Special Relativity of a symmetric
2-tensor $T$ representing energy, stresses and momentum densities which
satisfy conservation laws was a motivation for Einstein in the choice%
\footnote{%
With the help of his friend the mathematcian Grossman.} of its non vacuum
equations 
\begin{equation}
S_{\alpha \beta }=\kappa T_{\alpha \beta },\text{ }
\end{equation}
with $\kappa $ a constant usually normalized to $1$ by mathematicians. The
problem is then the resolution of the coupled system of Einstein with
sources and the conservation laws for these sources, 
\begin{equation}
\nabla _{\alpha }T^{\alpha \beta }=0,
\end{equation}
with also eventually equations for fields other than gravitation, for
instance Maxwell equations in presence of an electromagnetic field.

The Cauchy problem \ for the Einstein equations with sources splits again as
constraints on initial data and an evolution problem for reduced Einstein
equations with sources. The treatment of the electrovacuum case is similar
to vacuum and was solved simultaneously\footnote{%
Y. Four\`{e}s-Bruhat ''\textit{Th\'{e}or\`{e}me d'existence et d'unicit\'{e}
dans les th\'{e}ories relativistes de l'\'{e}lectromagn\'{e}tisme''} C.R.
Acad.Sci. \textbf{232} 1951}. Solution in the cases of dust and perfect
fluids without or with charge and zero conductivity were shown to admit a
well posed Cauchy problem\footnote{%
Y. Four\`{e}s-Bruhat ''\textit{Theor\`{e}mes d'existence en m\'{e}canique
des fuides relativistes''} \ Bull. Soc. France \textbf{86}, 155-175, 1958.}
using the Leray theory of hyperbolic systems\footnote{%
J.\ Leray \textit{''hyperbolic differential equations''\ }\ Mimeographed
Notes IAS, 1953}; relativistic fluids with infinite conductivity were
analysed\footnote{%
Y.\ Bruhat \textit{Fluides relativistes de conductivity infinie}
Astronautica Acta \textbf{VI}, 354-365, 1961.}. They were seen to be what is
now called Leray-Ohya\footnote{%
J. Leray and Y. Ohya Math Annalen \textbf{162}, 228-236 1968.
\par
It was shown later by K.O. Friedrichs using general Lagrangian methods that
fluids with infinite conductivity satisfy a first order symmetric hyperbolic
system. See for instance A.\ M. Anile Relativistic fluids and magnetofluids
\ Cambridge University press 1989} hyperbolic when Leray and Ohya proved
well posedness of the Cauchy problem in Gevrey classes for some systems of
differential equations with multiple characteristics. All these results
obtained for barotropic fluids were extended by Lichnerowicz to fluids whose
equation of state depends also on the entropy\footnote{%
As suggested by A.Taub, on physical grounds.} and assembled by him into a
book\footnote{%
A. Lichnerowicz \textit{Relativistic fluids and magneto fluids''} Benjamin
1967.} after a series of lectures he gave in Dallas at the invitation of
Ivor Robinson. Relativistic fluids with finite conductivity were proved to
be also Leray-Ohya hyperbolic\footnote{%
Y \ Choquet-Bruhat \textit{''Etude des \'{e}quations des fluides charg\'{e}s
relativistes inductifs et conducteurs''} Comm. Math. Phys. \textbf{3}
334-357\ 1966}. Isotropic relativistic elasticity has been proved by Pichon
to obey a Leray-Ohya hyperbolic system\footnote{%
G.\ Pichon \textit{\ ''Th\'{e}or\`{e}mes d'existence pour les \'{e}quations
des milieux \'{e}lastiques''} J. Math. Pures. et App \ \textbf{45} 3395-409
1966}. The hyperbolic character, Leray or Leray Ohya, holds for the Einstein
equations coupled with any of the quoted sources. The equations of charged
fluids with electromagnetic inductions are also Leray- Ohya hyperbolic, but
their natural Maxwell tensor being non symmetric their coupling with
Einstein equations is problematic\footnote{%
See M. Q. Pham \textit{Etude \'{e}lectrodynamique et thermodynamique dun
fluide relativiste charg\'{e} J.\ Rat. Mech. Anal. }\textbf{5}, 473-538
,1956. Various symmetrizations have been proposed along the years, but their
conservation laws lead to very unpleasant equations with unphysical
interpretations. The physical answer -seems to be that at the scale where
inductions play a role the gravational field is negligible.}.

Well posedness of the Cauchy problem was proved to be true for sources
satisfying a Vlasov\footnote{%
Case of particles with a given rest mass and no charge:
\par
Y. Choquet-Bruhat ''\textit{Solution du probl\`{e}me de Cauchy pour le
syst\`{e}me int\'{e}gro-differentiel d'Einstein Liouville''} \ Ann..Inst.
Fourier \textbf{XXI 3} 181-203 1971.
\par
Case of particles with electric charge:and arbitrary (positive) rest masses
using a weight factor to obtain convergent integrals:
\par
Y. Choquet-Bruhat ''\textit{Existence and uniqueness for the
Einstein-Maxwell- Liouville'' system'' }Volume in honor of Professor
Petrov,60th birthday Kiev\textit{\ 1971.}}, or a Boltzman\footnote{%
D. Bancel \textit{''Probl\`{e}me de Cauchy pour l'\'{e}quation de Boltzman
en Relativit\'{e} G\'{e}n\'{e}rale''} Ann. Inst. Poincar\'{e} XVIII n$^{0}$\
3 263-284 1971
\par
D. Bancel and Y. Choquet-Bruhat \textit{''Existence, uniqueness and local
stability for the Einstein-Botzman system''} Com. Math. Phys.1-14 1973.}
equation with appropriate cross section.

\section{Constraints}

We said that it has long been known that geometric initial data for the
vacuum Einstein equations on a spacelike submanifold $M$ are the two
fundamental forms, induced metric $\bar{g}$ and extrinsic curvature $K,$ and
they must satisfy $n$ equations, the constraints. Surprisingly it took a
long time to split these data into arbitrarily given quantities and unknowns
which satisfy elliptic equations, as it was however reasonable to expect for
unknowns on a space manifold and the Newtonian approximation of the Einstein
equations. The first result in this direction was due to Racine\footnote{%
Ch. Racine ''\textit{Le probl\`{e}me des }$n$\textit{\ corps dans la
th\'{e}orie de la Relativit}\'{e} ''\ Th\`{e}se Paris 1934, Gauthier
Villars.\ }. He assumed, for $n+1=4,$ the metric $\bar{g}$ to be conformally
flat 
\begin{equation*}
\bar{g}:=\phi ^{4}e,\text{ \ \ \ \ \ }e\text{\ the Euclidean metric}
\end{equation*}
and remarked that, if the trace $\bar{g}^{ij}K_{ij}$ of the extrinsic
curvature $K$ vanishes and one sets 
\begin{equation*}
P_{ij}=\phi ^{2}K_{ij},
\end{equation*}
the system of constraints for the equations with source of zero momentum
splits into a first order linear system for $P$, independent of $\phi ,$ and
a semi linear second order equation for $\phi $ with principal term the
Laplacian $\Delta \phi .$

The study was taken anew by Lichnerowicz\footnote{%
A. Lichnerowicz ''L'int\'{e}gration des \'{e}quations de la gravitation
relativiste et le probl\`{e}me des $n$ corps; J. Math. pures et App. 37-63,
1944.}, replacing the Euclidean metric by a general Riemannian metric $%
\gamma .$ He defines the traceless tensor $\tilde{K}_{ij}$ by 
\begin{equation}
\tilde{K}_{ij}=\varphi ^{2}(K_{ij}-\frac{1}{3}\bar{g}_{ij}\tau ),\text{ \ \
\ }\tau :=\bar{g}^{ij}K_{ij}.
\end{equation}
The momentum constraint reads then as the linear system for $\tilde{K}$ 
\begin{equation}
D_{i}\tilde{K}^{ij}=\frac{2}{3}\varphi ^{6}\gamma ^{ij}\partial _{i}\tau
+\varphi ^{10}J^{j},
\end{equation}
independent of $\varphi $ if the initial surface is maxima (he says
''minima'') i.e. $\tau =0,$ and if the momentum $J$ of the sources is zero.
The Hamiltonian constraint reads then as a second order elliptic equation
with only unknown $\varphi $ when $\tilde{K}$ and the matter density $\rho $
are known 
\begin{equation}
8\Delta _{\gamma }\varphi -R(\gamma )\varphi +|\tilde{K}|_{\gamma
}^{2}\varphi ^{-7}+(\rho -\frac{2}{3}\tau ^{2})\varphi ^{5}=0.\text{ \ }
\end{equation}

Lichnerowicz constructs a class of exact initially static data for the $N$
body problem with supports in domains $D_{I},$ $I=1,...n$, and matter
densities $\mu _{I}$ by taking $\gamma =e,$ assuming $J=0,\tau =0$ and
taking $\tilde{K}=0$ as a solution of the momentum constraint. The system of
constraints reduces then to the elliptic non linear equation with pricipal
term the Euclidean Laplace operator 
\begin{equation*}
\Delta \varphi =f(\varphi )\text{ \ \ with }f(\varphi )=0\text{\ \ in vacuum
and \ }f(\varphi )\equiv -\frac{1}{8}\mu _{I}\varphi ^{5}\text{ \ in \ }%
D_{I}.
\end{equation*}
Using the potential formula Lichnerowicz solves this equation by iteration,
showing the convergence of the series for small enough $\mu _{I}.$ The
problem of meaningful non static solutions of the momentum constraint
remained unsolved.

It is only in 1961, writing an article on the Cauchy problem, for the book
edited by Louis Witten\footnote{%
Y. Bruhat ''\textit{The cauchy problem''} in ''\textit{Gravitation, an
introduction to current research}'', L. Witten ed Wiley 1962} and inspired
by a paper of D. Sharp\footnote{%
D.\ Sharp ''\textit{One and two surfaces formulation of the boundary value
problem for the Einstein- Maxwell equations''} thesis Princeton University
1961} on possible constraints for the arbitrary quantities of the ''thin
sandwich conjecture''\ of J. A. Wheeler, namely lapse and shift, which did
not lead to an elliptic system, that I realized that such an elliptic system
can be written for the corresponding spacetime densities $\mathcal{G}^{00}$
and $\mathcal{G}^{i0}$ by using the splitting of the Einstein equations
obtained through the harmonic gauge. A. Vaillant-Simon\footnote{%
A. Vaillant-Simon, J. maths pures et App\textbf{\ 48}; 1-90, 1969.}
constructed a solution of this system near from the Minkowski spacetime.

In 1971 I wrote an elliptic, but not quasidiagonal\footnote{%
Y. Choquet-Bruhat Com. Math. Phys. \textbf{21 }211-218,\textbf{\ }1971.},
system for geometric data on an arbitrary spacelike manifold which
stimulated the interest of J. York, then a student of J. A. Wheeler, in the
constraint problem. York remarked that the assumption ''maximal'' on the
initial manifold made for conformally formulated constraints can be replaced
by constant mean extrinsic curvature\footnote{%
J. W. York ''\textit{Role of conformal 3 geometry in the dynamics o
gravitation'' } Phys. Rev. lett. \textbf{28} 1082.1972.
\par
The constant non zero situation was neglected by previous authors who were
only interested in the asymptotically Euclidean manifolds where it does not
occur.} and he introduced weights for the sources $\rho $ and $J,$
physically justified at least for electromagnetic sources and dimension $%
n=3. $ He thus obtained the linear momentum constraint independent of $%
\varphi $%
\begin{equation*}
D_{i}\tilde{K}^{ij}=\tilde{J}^{i}.
\end{equation*}
The Hamiltonian constraint becomes then the nonlinear elliptic equation with
only unknown $\varphi $ when $\gamma $ is chosen, $\tilde{K}$ computed and $%
\tilde{\rho}$ is known 
\begin{equation*}
\text{\ }8\Delta _{\gamma }\varphi -R(\gamma )\varphi +|\tilde{K}|_{\gamma
}^{2}\varphi ^{-7}+\tilde{\rho}\varphi ^{-3}-\frac{2}{3}\tau ^{2}\varphi
^{5}=0.\text{\ }
\end{equation*}

A decomposition theorem for symmetric 2-tensors, linked to the fact that Lie
derivatives of vector fields span the $L^{2}$ dual of divergence free
symmetric 2-tensors, had been known for some time. It leads to the writing%
\footnote{%
See J\ W. York ''\textit{Decomposition of symmetric 2-tensors in the theory
of gravitation}''\ Annales de l'IHP A \textbf{4}, 319-331 1974 \ and
references therein.} of the general solution of the momentum constraint,
when $\tau $ is constant, under the form, with $U$ an arbitrary given
traceless symmetric 2 tensor

\begin{equation*}
\tilde{K}^{ij}=(\mathcal{L}_{\gamma ,conf}X)^{ij}+U^{ij}+\frac{1}{3}\gamma
^{ij}\tau ,\text{ \ \ \ }(\mathcal{L}_{\gamma
,conf}X)^{ij}:=D^{i}X^{j}+D^{j}X^{i}-\frac{2}{3}\gamma ^{ij}D_{k}X^{k}.
\end{equation*}
The vector field $X$ is then solution of the linear system 
\begin{equation*}
\Delta _{\gamma ,conf}X:=D.(\mathcal{L}_{\gamma ,conf}X)=-D.U+\tilde{J}.
\end{equation*}
which can be shown to be equivalent to a linear elliptic second order
operator for $X,$ to which known theorems can be applied.

The conformal method gave to the Hamiltonian constraint on a manifold a
geometric comparatively simple form but non linear with no known generic
solution. I thought of applying to it the Leray-Schauder degree theory. I
brought to Leray in 1962 \ a Note about its solution in H\"{o}lder spaces
for publication in the C.R. of the french Academy of sciences. Leray
remarked that my result would hold for more general equations, and suggested
we publish jointly the general result. It was for me a great honor. Leray
wrote\footnote{%
Y Choquet-Bruhat et J. Leray ''\textit{Sur le probl\`{e}me de Dirichlet
quasi-lin\'{e}aire d'ordre 2''} C.\ R. Acad. Sci \textbf{274} 81-85 1972}
very fast for compact manifolds this Note which introduces sub and super
solutions, and refused to cosign the Note solving the particular case of the
Hamiltonian constraint which I wrote shortly afterwards\footnote{%
Y. Choquet-Bruhat ''S\textit{olution globale du probl\`{e}me des contraintes
sur une vari\'{e}t\'{e} compacte''} C.R. Acad Sci. \textbf{274} 682-684 1972.%
}. I obtained later the result for asymptotically Euclidean manifolds in
weighted H\"{o}lder spaces\footnote{%
Y. Choquet-Bruhat ''\textit{Solution of the problem of constraints on open
and closed manifolds}'' J. Gen. Rel and Grav.\textbf{\ 5} 45-64 1974.}. A
large amount of work has been devoted since that time to the solution of the
constraints expressed as the elliptic semilinear system obtained by the
conformal method on a constant mean curvature initial manifold, and using
weighted Sobolev spaces. Progress has been made in lifting the constant mean
curvature hypothesis and weakening the regularity, but there is space for
further work.

\section{Local existence and global uniqueness.}

In the beginning of the seventies the geometric character of the Cauchy
problem for the Einstein equations was well understood\footnote{%
Let us quote in addition to previously quoted papers:
\par
J.\ A. Wheeler in ''Relativity, Gorups and Topoloy'', B. and C. DeWitt ed.
Gordon and Breach 1964
\par
R. Penrose Phys Rev. lett. 14 57 1965,
\par
S. Hawking Proc.Roy. Soc \textbf{294}\ A\ 511, 1966}. The global in space,
local in time, existence was known for the classical sources mentioned
above, at least for compact or asymptotically Euclidean manifolds, apart
from lessening the regularity required of data and abandoning the constant
mean curvature hypothesis of the initial manifold. Local uniqueness, up to
diffeomorphisms, of a solution of the evolution of geometric data satisfying
the constraints was also known, but the question of a global isomorphism
between solutions was open. In fact, though known of specialists, the
geometric local existence and uniqueness did not have in the litterature a
concise and precise formulation. I convinced Geroch we write up proper
definitions for the geometric local theorem\footnote{%
Y. Choquet-Bruhat and R. Geroch ''\textit{Probl\`{e}me de Cauchy
intrins\`{e}que en Relativit\'{e} G\'{e}n\'{e}rale}'' C\ . R. Acad. Sci. A 
\textbf{269} 746-748, 1969.} before publishing our global geometric
uniqueness proof.

A fundamental notion for the study of global solutions of linear hyperbolic
systems of arbitrary order had been introduced in 1952 by J. Leray. He had
defined what he called ''time like paths''\ for such systems and defined
global hyperbolicity as compactness (in the space of paths) of any, non
vacuum, set of timelike paths joining two arbitrary points. This general
definition applies in particular to Lorentzian manifolds. In this case
global hyperbolicity was proved (Choquet-Bruhat 1967) to be equivalent to
the strong causality defined by Penrose 1967 added to the compactness of the
spacetime domaines defined by intersections of past and future of any two
spacetime points (see definitions and proof in the Hawking and Ellis book of
1973).

In 1969, the local existence was completed\footnote{%
Y.\ Choquet-Bruhat and R. Geroch ''\textit{Global aspects of the Cauchy
problem in General Relativity}'' Comm. Math. Phys. \textbf{14}, 329-335,
1969.} by a geometric global uniqueness result, namely existence and
uniqueness, , up to isometries, in the class of globally hyperbolic
spacetimes, of a maximal\footnote{%
i.e. which cannot be isometrically embedded into a larger Einsteinian
spacetime.} Einsteinian development of given initial data. Geroch and myself
had met and discussed at the ''Batelle rencontres 1967''\ organized by J. A.
Wheeler and C. DeWitt. We obtained a complete proof during a visit we both
made in London\footnote{%
I had obtained in 1968 a global uniqueness theorem, but restricted to
complete Einsteinian spacetimes.}. In 1970 Geroch proved the following very
useful criterium: global hyperbolicity is equivalent to the existence of a
''Cauchy surface'', 3 manifold cut once and only once by any timelike curve
without end point.

The geometric global uniqueness result, first proved in the vacuum case,
extends to Einstein equations with sources which have a well posed causal
Cauchy problem; that is, in particular, satisfy Leray or Leray-Ohya
hyperbolic systems.

\section{Global existence and singulariries.}

Generic problems of global existence or formation of singularities were,
towards the end of the sixties, mainly open. They became for analysts and
geometers interested in the modeling of the world we live in, a vast field
of research. It led to new definitions, conjectures, remarkable results, and
new open problems.

\begin{center}
\textbf{References\footnote{%
Y Bruhat, Y. Choquet-Bruhat and Y.\ Four\`{e}s-Bruhat are the same person,
daughter of the physicist Georges Bruhat..}}
\end{center}

D. Bancel \textit{''Probl\`{e}me de Cauchy pour l'\'{e}quation de Boltzman
en Relativit\'{e} G\'{e}n\'{e}rale''} Ann. Inst. Poincar\'{e} XVIII n$^{0}$\
3 263-284 1973

D. Bancel and Y. Choquet-Bruhat \textit{''Existence, uniqueness and local
stability for the Einstein-Botzman system''} Com. Math. Phys.1-14 1973.

Y. Bruhat ''\textit{Fluides relativistes de conductivit\'{e} infinie''}
Astronautica Acta 6 354-355 1961

Y. Bruhat ''\textit{The Cauchy problem}'' in ''\textit{Gravitation, an
introduction to current research''} Louis Witten ed. Wiley 1962

Y.\ Bruhat \textit{''Probl\`{e}me des conditions initiales sur un
cono\"{\i}de caract\'{e}ristique} C.R.\ Acad. Sci \textbf{256}, 371-373,
1963.

Y. Bruhat ''\textit{Sur la th\'{e}orie des propagateurs}'' Annali di
Matematica Serie IV, tomo LXIV -1964.

F. Cagnac ''\textit{Probl\`{e}me de Cauchy sur les hypersurfaces
caract\'{e}ristiques des \'{e}quations d'Einstein du vide''} C.R. Acad Sci 
\textbf{262} 1966

F. Cagnac ''\textit{Probl\`{e}me de Cauchy sur un cono\"{i}de
caract\'{e}ristique pour les \'{e}quations d'Einstein du vide''} Annali di
matemetica pura e applicata 1975.

Y.\ Choquet-Bruhat ''\textit{Etude des \'{e}quations des fluides charg\'{e}s
relativistes inductifs et conducteurs}'' Com. Math. Phys \textbf{3} 334-357
1966.

Y.\ Choquet-Bruhat ''Hyperbolic partial differential equations on a
manifold'' in C.\ DeWitt and J.A. Wheeler ed \textit{''Batelle rencontres
1967 in Mathematics and Physics''}

Y.\ Choquet-Bruhat ''\textit{Espaces Einsteiniens g\'{e}n\'{e}raux, chocs
gravitationnels}'' Ann. Inst. Poincar\'{e}, 8 n$%
{{}^\circ}%
4$, 327-338, 1968.

Y.\ Choquet-Bruhat ''\textit{Un th\'{e}or\`{e}me global d'unicit\'{e} pour
les solutions des \'{e}quations d'Einstein}'' Bul. Soc. Math. France \textbf{%
96 }181-192, 1968.

Y. Choquet-Bruhat \textit{''New elliptic system and global solutions for the
constraints equations in General Relativity''} Com. Math. Phys. \textbf{21 }%
211-218,\textbf{\ }1971.

Y. Choquet-Bruhat ''S\textit{olution globale du probl\`{e}me des contraintes
sur une vari\'{e}t\'{e} compacte''} C.R. Acad Sci. \textbf{274} 682-684 1972.

Y. Choquet-Bruhat ''\textit{Solution of the problem of constraints on open
and closed manifolds}'' J. Gen. Rel and Grav.\textbf{\ 5} 45-64 1974.

Y.\ Choquet-Bruhat and R. Geroch ''\textit{Probleme de Cauchy
intrins\`{e}que en Relativit\'{e} g\'{e}n\'{e}rale}'' C. R. Acad. Sci. 269
746-748, 1969.

Y.\ Choquet-Bruhat and R. Geroch ''\textit{Global aspects of the Cauchy
problem in General Relativity}'' Comm. Math. Phys. \textbf{14}, 329-335,
1969.

Y Choquet-Bruhat et J. Leray ''\textit{Sur le probl\`{e}me de Dirichlet
quasi-lin\'{e}aire d'ordre 2''} C\ R. Acad. Sci \textbf{274} 81-85 1972

G. Darmois ''\textit{Les equations de la gravitation Einsteinienne}''
Memorial des Sciences math\'{e}matiques Gauthier Villars 1927.

B. DeWitt.\textit{''Quantization of geometry''} in \textit{Les Houches 1963}
Gordon and Breach

De Donder \textit{''La gravifique Einsteinienne''} Memorial des Sciences
math\'{e}matiques Gauthier Villars 1925.

Y. Four\`{e}s-Bruhat ''\textit{Th\'{e}or\`{e}mes d'existence pour certains
syst\`{e}mes d'\'{e}quations aux d\'{e}riv\'{e}es partielles non
lin\'{e}aires}'' Acta Mathematica \textbf{88, 42-225,}\ 1952.

Y. Four\`{e}s-Bruhat ''\textit{Th\'{e}or\`{e}me d'existence et d'unicit\'{e}
dans les th\'{e}ories relativistes de l'\'{e}lectromagn\'{e}tisme''} C.R.
Acad.Sci. \textbf{232} 1951.

Y. Four\`{e}s-Bruhat ''\textit{R\'{e}solution du probl\`{e}me de Cauchy pour
des \'{e}quations hyperboliques du second ordre non lin\'{e}aires''} Bull.
Soc. Math. France \textbf{81}, 225-288 1953

Y. Four\`{e}s-Bruhat \textit{''Solution \'{e}l\'{e}mentaire d'\'{e}quations
ultrahyperboliques''} J. Math. Pures et Ap \textbf{121} 277-289 1955.

Y. Four\`{e}s-Bruhat ''\textit{sur l'int\'{e}gration des \'{e}quations de la
Relativit\'{e} G\'{e}n\'{e}rale}'' J. Rat. Mech. and Anal. \textbf{5}
951-966 1956.

Y. Four\`{e}s-Bruhat ''\textit{Theor\`{e}mes d'existence en m\'{e}canique
des fluides relativistes''} \ Bull. Soc. France \textbf{86}, 155-175, 1958.

R. Geroch ''\textit{The domain of dependence}''\ J. Math. Phys \textbf{11}
437-439 1970

S.\ W. Hawking and G.R.F Ellis \textit{''The large scale structure of
spacetime''} Cambridge university press 1973.

J. Leray \textit{''Hyperbolic differential systems''} Mimeographed Notes,
IAS Princeton 1953.

J. Leray and Y. Ohya Math Annalen \textbf{162}, 228-236 1968.

A. Lichnerowicz ''\textit{Probl\`{e}mes globaux en m\'{e}canique relativiste}%
'', Hermann 1939.

A. Lichnerowicz. ''\textit{L'int\'{e}gration des \'{e}quations de la
gravitation relativiste et le probl\`{e}me des }$n$\textit{\ corps''} J.
Math. pures et App. 37-63, 1944.

A. Lichnerowicz \textit{'' Th\'{e}ories relativistes de la gravitation et de
l'\'{e}lectromagn\'{e}tisme''} Masson, Paris 1955.

A. Lichnerowicz ''\textit{Propagateurs et commutateurs en Relatitit\'{e}
G\'{e}n\'{e}rale'', }publications math\'{e}matiques de l'IHES, 1$,$ 1961%
\textit{\ }

N. O'Murchada and J. W. York ''\textit{The initial value problem of General
Relativity''} Phys. Rev. D \textbf{10} 428-446 1974.

R.\ Penrose \textit{''structure of spacetime''} in C.\ DeWitt and J.A.
Wheeler ed \textit{''Batelle rencontres 1967 in Mathematics and Physics''}

M. Q. Pham \textit{Etude \'{e}lectrodynamique et thermodynamique d'un fluide
relativiste charg\'{e} J.\ Rat. Mech. Anal. }\textbf{5}, 473-538 ,1956.

G.\ Pichon. ''\textit{Etude relativiste de fluides visqueux et charg\'{e}s}%
'' Ann. Inst. Poincar\'{e} II n$^{0}1$ 1965.

G.\ Pichon \textit{\ ''Th\'{e}or\`{e}mes d'existence pour les \'{e}quations
des milieux \'{e}lastiques''} J. Math. Pures. et App \ \textbf{45} 3395-409
1966

Ch. Racine C.R. Acad Sci. Paris 1931.

Ch. Racine ''\textit{Le probl\`{e}me des }$n$\textit{\ corps dans la
th\'{e}orie de la Relativit\'{e} }'' Th\`{e}se Paris 1934, Gauthier
Villars.\ 

J. Schauder Fundamenta mathematicae, 24 213-246 1935.

D.\ Sharp ''\textit{One and two surfaces formulation of the boundary value
problem for the Einstein- Maxwell equations''} thesis Princeton University
1961

S.\ Sobolev \ \textit{''Methode nouvelle \`{a} r\'{e}soudre le probl\`{e}me
de Cauchy pour les \'{e}quations lin\'{e}aires hyperboliques normales''}
Rec. Math. Moscou N. s. 1936.

K. Stellmacher. Math. Annalen 115, 1938.

A.H. Taub Phys. Rev \textbf{74 } 328-334 1948

A.H. Taub J. Rat. Mech. Anal. \textbf{3} 312-324 1959.

A. Vaillant-Simon, J. maths pures et App\textbf{\ 48}; 1-90, 1969.

J.\ A. Wheeler in ''Relativity, Groups and Topoloy'', B. and C. DeWitt ed.
Gordon and Breach 1964

J. W. York ''\textit{Role of conformal 3 geometry in the dynamics of
gravitation'' } Phys. Rev. lett. \textbf{28} 1082.1972.

J\ W. York ''\textit{Decomposition of symmetric 2-tensors in the theory of
gravitation}'' Annales de l'IHP A \textbf{4}, 319-331 1974.

\end{document}